\documentclass[aps,prd,superscriptaddress,long,notitlepage,balancelastpage,nofootinbib,floatfix,twocolumn]{revtex4-1}
\pdfoutput=1
\usepackage{amsmath,mathtools,amssymb,amsthm,amsxtra,overpic,bbm,epsfig,subfigure,url,bm}
\usepackage{hyperref}
\usepackage{mathrsfs}
\usepackage{color,xcolor}
\usepackage{comment}
\usepackage{float}
\usepackage{enumitem}
\usepackage{slashed}
\usepackage{multirow}
\usepackage{appendix}
\usepackage[normalem]{ulem}
\DeclareUnicodeCharacter{03BC}{\ensuremath{\mu}}
\DeclareUnicodeCharacter{2032}{\ensuremath{'}}
\definecolor{nicered}{rgb}{0.5,0.,0.}
\definecolor{nicegreen}{rgb}{0.,0.5,0.}
\definecolor{niceblue}{rgb}{0.,0.,0.5}
\hypersetup{
	colorlinks=true,
	linkcolor=black,
	filecolor=nicegreen,      
	urlcolor=niceblue,
	citecolor=nicered,
}

\makeatletter
\newcommand*{\balancecolsandclearpage}{%
	\close@column@grid
	\cleardoublepage
	\twocolumngrid
}

\newcommand{\beq}{\begin{equation}}
\newcommand{\eeq}{\end{equation}}
\newcommand{\bea}{\begin{eqnarray}}
\newcommand{\ena}{\end{eqnarray}}

\def \epsilon {\varepsilon} 

\def \vec#1{{\boldsymbol{#1}}}

\DeclareUnicodeCharacter{2212}{\textendash}

\makeatother  
\begin{document}

\title{\vspace{1cm} \large 
Long-lived doubly charged scalars in the left-right symmetric model: \\ catalyzed nuclear fusion and collider implications
}

\author{\bf Evgeny Akhmedov}
\email[E-mail:]{akhmedov@mpi-hd.mpg.de}
\affiliation{Max-Planck-Institut f{\"u}r Kernphysik, Saupfercheckweg 1, 69117 Heidelberg, Germany}

\author{\bf P. S. Bhupal Dev}
\email[E-mail:]{bdev@wustl.edu}
\affiliation{Department of Physics and McDonnell Center for the Space Sciences, Washington University, St.~Louis, Missouri 63130, USA}

\author{\bf Sudip Jana}
\email[E-mail:]{sudip.jana@mpi-hd.mpg.de}
\affiliation{Max-Planck-Institut f{\"u}r Kernphysik, Saupfercheckweg 1, 69117 Heidelberg, Germany}

\author{\bf Rabindra N. Mohapatra}
\email[E-mail:]{rmohapat@umd.edu}
\affiliation{Maryland Center for Fundamental Physics and Department of Physics, University of Maryland,
College Park, Maryland 20742, USA}

\begin{abstract}
We show that the doubly charged scalar from the $SU(2)_R$-triplet Higgs field in the Left-Right Symmetric Model has its mass governed by a hidden symmetry so that its value can be much lower than the $SU(2)_R$ breaking scale. This makes it a long-lived particle while being consistent with all existing theoretical and experimental constraints. Such long-lived doubly charged scalars have the potential to trigger catalyzed fusion processes in light nuclei, which may have important applications for energy production. We show that it could also bear consequences on the excess of large ionization energy loss ($dE/dx$) recently observed in collider experiments.
\noindent 
\end{abstract}
\maketitle
\section{Introduction} \label{sec:I}
Since the discovery of the Standard Model (SM) Higgs boson~\cite{ATLAS:2012yve, CMS:2012qbp}, no new particles have emerged from the Large Hadron Collider (LHC). This might suggest that the current energy and luminosity of the LHC may not be sufficient for direct exploration of beyond the Standard Model (BSM) particles. Optimism for potential discoveries relies on the high-luminosity LHC upgrade or higher-energy colliders. An alternative possibility, however, is that there are long-lived particles (LLPs) that may yield null results in mainstream prompt signal searches at collider experiments and yet are part of new physics. There is certainly mounting interest (both theoretical and experimental) in unconventional investigative approaches in the search for the LLPs~\cite{Fairbairn:2006gg, Alekhin:2015byh, Curtin:2018mvb, Lee:2018pag, Alimena:2019zri, Feng:2022inv, Knapen:2022afb}. In this context, we note that most of the LLP searches focus on neutral LLPs, being theoretically motivated by feebly-interacting neutral BSM particles like dark matter, axions, heavy neutral leptons, and new neutral gauge bosons, whereas charged LLPs have received limited attention so far. It is worth remembering that the 1930s' discovery of muons, unexpected charged long-lived particles observed by Anderson and Neddermeyer~\cite{Anderson:1936zz, Neddermeyer:1937md}, underscores the importance of vigilance, as breakthroughs may emerge in unforeseen ways. In this letter, we scrutinize the prospect that the doubly charged scalars inherent to the minimal left-right symmetric model (LRSM) could embody the distinctive characteristics of such LLPs.

Doubly charged scalars emerge in various BSM scenarios, such as the type-II seesaw model~\cite{Magg:1980ut, Schechter:1980gr, Lazarides:1980nt, Mohapatra:1980yp}, left-right symmetric model~\cite{Pati:1974yy, Mohapatra:1974gc, Senjanovic:1975rk, Mohapatra:1979ia}, radiative neutrino mass model~\cite{Zee:1985id, Babu:1988ki}, little Higgs model~\cite{Arkani-Hamed:2002iiv}, $d=7$ neutrino mass models~\cite{Babu:2009aq, Bonnet:2009ej}, $331$ model~\cite{Pisano:1992bxx}  and the Georgi-Machacek Model~\cite{Georgi:1985nv}. In most instances, they manifest themselves either as $SU(2)_L$ singlets with hypercharge $Y=2$, as seen in loop models for neutrino masses, or as members of a triplet of $SU(2)_L$, as in the type-II seesaw models. In the LRSM, they can exist as either $SU(2)_L$-singlet or triplet. However, they may also arise from a doublet, quadruplet, or quintuplet of $SU(2)_L$~\cite{Babu:2009aq, Bonnet:2009ej, Babu:2019mfe, Babu:2020hun, Bhattacharya:2016qsg}. If doubly charged scalars originate from $SU(2)_L$ multiplets other than singlets, the challenge of conferring longevity becomes particularly intricate, primarily due to potential interactions with two same-sign $W$ bosons. Additionally, the masses of charged or neutral partners within the $SU(2)_L$ multiplet of doubly charged Higgs are tightly constrained by electroweak $\rho$ parameter contributions, indicating anticipated signals at colliders. Designating doubly charged scalars as weak isosinglets classifies them aptly as LLP if their couplings to lepton pairs are either suppressed or set to zero by some symmetry. In general, these scalars exhibit couplings to two like-sign charged leptons with arbitrary flavor content, crucial for their role in generating neutrino masses~\cite{Zee:1985id, Babu:1988ki}. Prohibiting their coupling to leptons diminishes their role in neutrino mass generation. In this context, the unique standing of the doubly charged scalar $\Delta^{\pm\pm}_R$ within the LRSM, belonging to the $SU(2)_L$-singlet and $SU(2)_R$-triplet, distinguishes it from other conceivable scenarios. In this letter, we show that we can have a long-lived doubly charged scalar and generate neutrino masses simultaneously.

The mass of the $\Delta^{\pm\pm}_R$ in the LRSM is commonly expected to be at the $SU(2)_R$ breaking scale. However, we show that due to the existence of a hidden symmetry, its magnitude is decoupled from the $W_R$ scale at the tree level. In other words, even for very high $W_R$ mass, the $\Delta^{\pm\pm}_R$ mass can be made very low by choice of a symmetry-breaking parameter in the Higgs potential. If we include one loop contribution to the mass, a mild fine tuning is sufficient to keep the mass low. We then discuss under what conditions these doubly charged bosons can be very long-lived and find constraints on them from current observations. Furthermore, we demonstrate that although suppressing the leptonic couplings precludes the use of the type-I seesaw mechanism~\cite{Minkowski:1977sc, Mohapatra:1979ia, Yanagida:1979as, Gell-Mann:1979vob, Glashow:1979nm} to explain small neutrino masses, the inverse seesaw mechanism~\cite{Mohapatra:1986aw, Mohapatra:1986bd} can be responsible for neutrino mass generation. 

We subsequently analyze the consequential phenomenological implications of long-lived doubly charged scalars. Recent findings indicate that sufficiently long-lived doubly charged scalars can facilitate the fusion of light nuclei, bearing significant practical implications for energy production~\cite{Akhmedov:2021qmr}.  Investigating their longevity necessary for catalyzed fusion, we also analyze the collider implications of these almost stable particles, characterized by distinctive ionization energy loss ($dE/dx$) behavior~\cite{Jager:2018ecz, Altakach:2022hgn}. Notably, recent observations by the ATLAS collaboration indicated an excess in $dE/dx$ for LLP searches~\cite{ATLAS:2022pib}, although subsequent data have neither confirmed nor ruled it out thus far~\cite{ATLAS:2023zxo}. Nevertheless, these long-lived doubly charged particles offer a compelling explanation for the observed excess~\cite{Giudice:2022bpq}, making them promising candidates for future LLP discoveries based on ionization energy loss gradients.

\section{$\Delta^{\pm\pm}_R$ in the minimal Left-Right Symmetric Model} \label{sec:II}
Let us start with a brief review of the minimal LRSM for neutrino masses~\cite{Pati:1974yy, Mohapatra:1974gc, Senjanovic:1975rk, Mohapatra:1979ia} and the uniqueness of the $\Delta^{\pm\pm}_R$ boson. The model is based on the gauge group $SU(3)_c \times SU(2)_L\times SU(2)_R\times U(1)_{B-L}$, with the quantum numbers of the fermion and Higgs fields given in Table~\ref{table:particle spectrum}. 
Here and in what follows, we omit the $SU(2)_L$-triplet field $\Delta_L$ $(1,3, 1, 2)$ since we will deal with the effective field theory at lower energies below the $D$-parity breaking scale and the $\Delta_L$ multiplet is expected to remain at the higher $D$-parity breaking scale in this version~\cite{Chang:1983fu}.
This ensures that the neutrino masses are given by the simple type-I seesaw formula linking the neutrino mass to the $W_R$ boson mass scale~\cite{Mohapatra:1979ia}, without any fine-tuning between the type-I and type-II seesaw contributions. Moreover, this version is more amenable to $SO(10)$ embedding with lighter $W_R$ bosons~\cite{Chang:1984qr}. Most importantly for our discussion here, as noted in Section~\ref{sec:I}, the doubly charged component $\Delta_L^{\pm\pm}$ cannot be long-lived because of its gauge coupling to $W$ bosons. Therefore, keeping it in the theory will only complicate the discussion without affecting our main results. Nevertheless, we analyze the full scalar potential, including the $\Delta_L$ field in Appendix~\ref{sec:app}.      

\begin{table}[!t]
\centering
\caption{Gauge quantum numbers of all the fields in the minimal LRSM based on $SU(3)_c\times SU(2)_L\times SU(2)_R\times U(1)_{B-L}$.}
\begin{tabular}{|c|c|c|}\hline
\multicolumn{2}{|c|}{Particles} & Quantum numbers \\ \hline
\multirow{8}{*}{fermions} & $Q_L = \begin{pmatrix}u_L\\d_L\end{pmatrix}$ & $(3, 2, 1, 1/3)$\\
& $Q_R = \begin{pmatrix}u_R\\d_R\end{pmatrix}$ & $(3, 1, 2, 1/3)$\\
& $\ell_L = \begin{pmatrix}\nu \\ e_L\end{pmatrix}$ & $(1, 2, 1, -1)$\\
& $\ell_R = \begin{pmatrix} N \\ e_R\end{pmatrix}$ & $(1, 1, 2, -1)$\\\hline
\multirow{4}{*}{scalars} & $\Phi=\begin{pmatrix} \phi_1^0 & \phi_2^+\\
\phi_1^- & \phi_2^0 
\end{pmatrix}$ &$(1,2,2,0)$\\
& $\Delta_R=\begin{pmatrix}
    \Delta_R^+/\sqrt{2} & \Delta_R^{++}\\
    \Delta_R^0 & -\Delta_R^+/\sqrt 2
\end{pmatrix}$ & $(1,1, 3,2)$\\
\hline
\end{tabular}\label{table:particle spectrum}
\end{table}

The Yukawa Lagrangian for the model  is given by
\begin{eqnarray}
{\cal L}_Y &=&  
h_q\bar{Q}_L \Phi Q_R ~+~ 
h'_q\bar{Q}_L \tilde{\Phi} Q_R ~+~ 
h_\ell \bar{\ell}_L \Phi \ell_R \nonumber \\
&& ~+~ 
h'_\ell \bar{\ell}_L \tilde\Phi \ell_R+
f_R{\ell}^T_R \Delta_R \ell_R ~+~ {\rm H.c.} \, ,
\label{eq:lag}
\end{eqnarray}
where $\tilde{\Phi}=\sigma_2\Phi^*\sigma_2$ (with $\sigma_2$ being the second Pauli matrix), and $h_{q,\ell}$, $h'_{q,\ell}$, $f_R$ are the Yukawa coupling matrices. 

In the scalar sector, after the neutral components of the bidoublet $\Phi$ and $SU(2)_R$-triplet $\Delta_R$ develop their vacuum expectation values (VEVs) [cf.~Eq.~\eqref{eq:vev}], there remain eight physical massive scalar fields, namely, four neutral ($h, H_1^0, A_1^0, H_3^0$) and pairs of singly-charged ($H^\pm$) and doubly charged ($H^{\pm\pm}\equiv \Delta_R^{\pm\pm}$) scalars; see Appendix~\ref{sec:app} for details. We identify $h$ as the SM-like Higgs boson with $m_h\simeq 125$ GeV [cf.~Eq.~\eqref{eq:h}]. The masses of the remaining scalar fields are proportional to the $\Delta_R$ VEV $v_R$~\cite{Dev:2016dja}: 
\begin{align}
    m_{H_1^0}^2 \simeq  m_{A_1^0}^2 \simeq m_{H^\pm}^2 \simeq & \ \alpha_3v_R^2 \, , \label{eq:H} \\
    m_{H_3^0}^2 \simeq & \ 4\rho_1v_R^2 \, , \label{eq:H3} \\
    m_{H^{\pm\pm}}^2 \simeq & \ 4\rho_2v_R^2 \, , \label{eq:Hpp} 
\end{align}
 The neutral Higgs masses  $m_{H_1^0, A_1^0}$ are constrained to be larger than about 15 TeV to satisfy the flavor-changing neutral current constraints from $K-\bar{K}$ and $B-\bar{B}$ transitions~\cite{Zhang:2007da, Maiezza:2010ic, Bertolini:2019out, Dekens:2021bro}. It follows from Eq.~\eqref{eq:H} that the singly-charged scalars $H^\pm$  must also be equally massive and are, therefore, inaccessible at the LHC~\cite{Dev:2016dja}. The $H_3^0$ field can in principle be much lighter (though at the cost of some fine-tuning), and may have interesting phenomenological consequences~\cite{Dev:2016nfr, Dev:2017dui, Dev:2023kzu}.    

As for the doubly charged scalars $H^{\pm\pm}$, an interesting point to note from Eq.~\eqref{eq:Hpp} is that their mass only depends on the quartic coupling $\rho_2$. This can be seen without doing any explicit calculation as follows: if  $\rho_2$  is set to zero in the potential~\eqref{eq:pot} at the tree level, it is not induced by loops. Therefore, we can tune down the value of $\rho_2$ arbitrarily. The presence of this naturally tunable parameter allows the mass of $H^{\pm\pm}$ to be not necessarily of the order of the $v_R$ scale so that it can be considerably lighter. The basic reason for this is that when $\rho_2$ and $\alpha_3$   are set to zero, 
the scalar potential has a higher global $SU(3)$ symmetry, not shared by gauge interactions, and as a result, the doubly charged boson is a pseudo-Goldstone particle with zero tree-level mass.  To see this, note that when $\langle\Delta^0_R\rangle = v_R\neq 0$, the above global $SU(3)$ symmetry breaks down to $SU(2)$ leaving $8-3=5$ Goldstone bosons. This VEV $v_R$ also breaks the gauge symmetry $SU(2)_R\times U(1)_{B-L}$ down to $U(1)_Y$. Out of the five massless scalar bosons, three are absorbed by the three massive gauge bosons of broken $SU(2)_R\times U(1)_{B-L}$ symmetry, leaving two massless states. These are the doubly charged states that pick up mass once the $\rho_2$ and $\alpha_3$ terms in the potential are switched on since
$\rho_2$ and $\alpha_3$ explicitly break this $SU(3)$ symmetry.
Thus $\rho_2$ and $\alpha_3$ play a crucial role in giving $H^{\pm\pm}$ mass in combination with $v_R$ and $\kappa$. Thus, the $H^{\pm\pm}$ could be considerably lighter than the $v_R$ scale since $\rho_2$ and $\alpha_3$ are symmetry-breaking parameters and could be adjusted to be small, although, in practice, $\alpha_3$ cannot be arbitrarily small in order to satisfy the FCNC constraints mentioned above. 
We note however that the one loop gauge boson mediated graphs will not respect the $SU(3)$ symmetry and will induce terms of the form $\frac{g^4}{32\pi^2}|{\rm Tr}(\Delta^\dagger_R\vec{\tau}\Delta_R)|^2\ln(v^2_R/\mu^2)$ to the potential. They can induce a mass for the $\Delta^{++}_R$ even in the limit of $\rho_2=0$ and $\alpha_3=0$. For high $v_R$, this induced  mass for $\Delta^{++}_R$, can be larger than a few TeV, that we prefer. However, we can cancel this contribution by a mild adjustment of $\rho_2$ and $\alpha_3$. For example for $v_R\sim 10^7$ GeV, a fine tuning at the level of $10^{-4}$ is sufficient. 

\section{Long-lived doubly charged scalars} \label{sec:III}
Our interest lies in the existence of a long-lived doubly charged scalar within a gauge theory. As mentioned in the introduction, a doubly charged scalar arising from $SU(2)_L$ singlet is crucial for achieving longevity. We show here that the $H^{\pm\pm}$ coming from the $SU(2)_R$-triplet fittingly qualifies to be long-lived.

\begin{figure}[htb!]
		\centering
		\includegraphics[width=0.49\textwidth, height=2.3 cm]{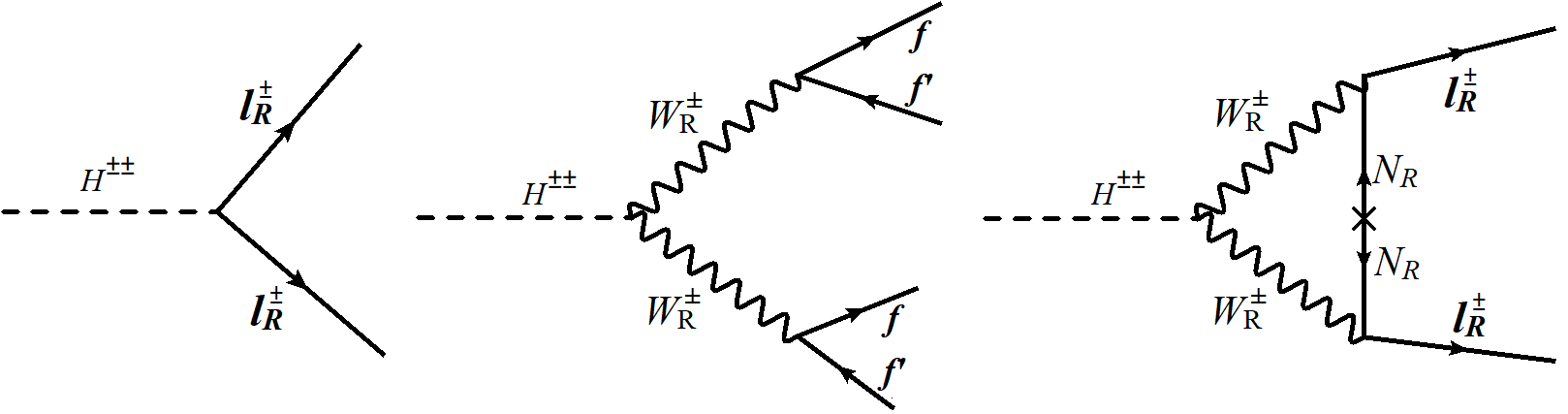}\\
 (a) \hspace{2cm} (b)  \hspace{2cm}  (c)
	\caption{Representative Feynman diagrams for dominant decay modes of $H^{\pm \pm}_R$. }
	\label{Fig:1}
\end{figure}
\begin{figure}[t!]
		\centering
		\includegraphics[width=0.49\textwidth]{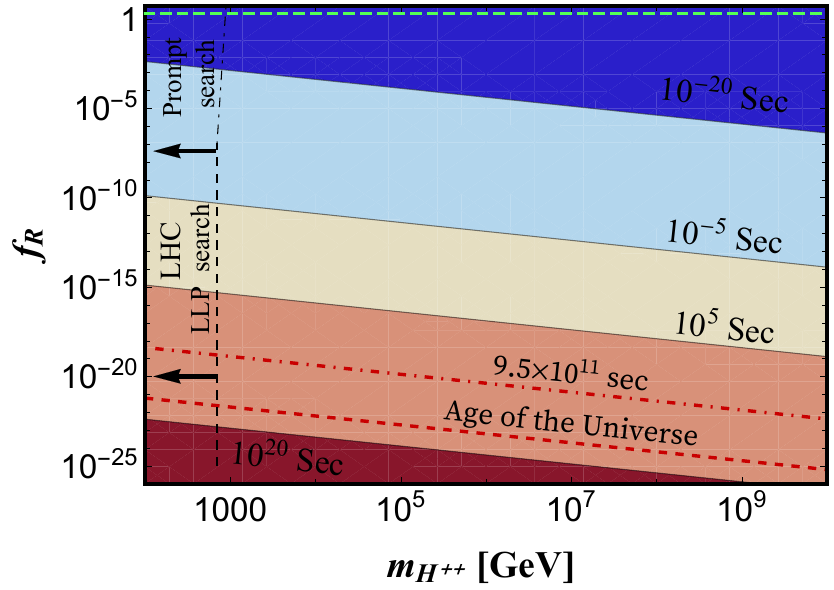}
	\caption{Lifetime of doubly charged scalars in the Yukawa coupling versus mass plane, with iso-contours representing different $H^{\pm\pm}$ lifetimes.  The almost vertical dashed lines show the current lower limits on the $H^{\pm\pm}$ mass from prompt and LLP searches at the LHC.  Red dot-dashed and dashed lines respectively correspond to two important lifetime benchmarks relevant to the catalyzed fusion mechanism: $\tau=9.5\times 10^{11}$ sec and $\tau=\tau_U=4.34\times 10^{17}$ sec (age of the Universe).  
  The cyan dashed line (near the top) indicates the perturbative limit on the Yukawa coupling. 
 }
	\label{Fig:2}
\end{figure} 
\begin{figure}[t!]
		\centering
		\includegraphics[width=0.49\textwidth]{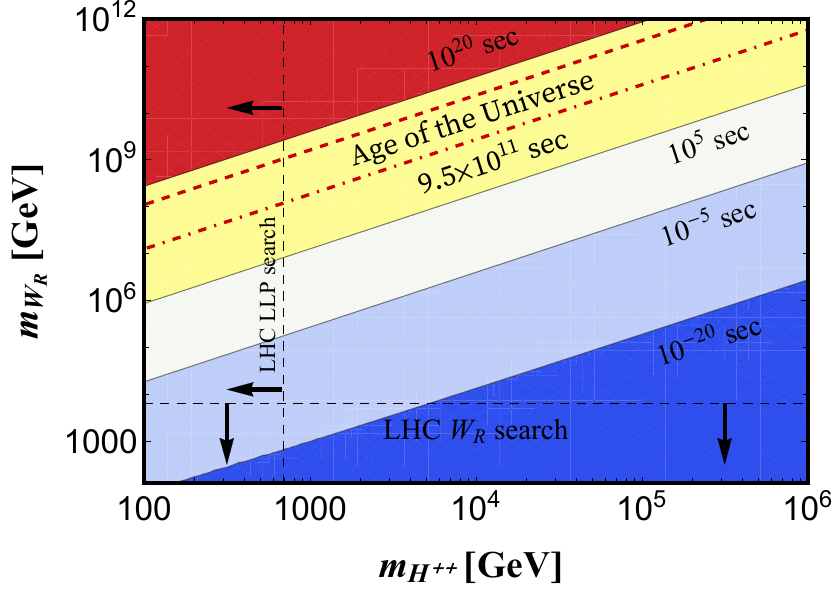}
	\caption{Lifetime of doubly charged scalars in the $W_R$ mass versus the $H^{\pm\pm}$ mass plane. Iso-contours denote different $H^{\pm\pm}$ lifetimes. Red dot-dashed and dashed lines correspond to  two important lifetime benchmarks relevant to the catalyzed fusion mechanism: $\tau=9.5\times
10^{11}$ sec and $\tau=\tau_U$. Black dashed lines indicate the lower limits on $H^{\pm\pm}$ and $W_R$ masses from  LLP searches and $W_R$ searches at the LHC, respectively. 
}
	\label{Fig:3}
\end{figure} 
The possible decay modes of $H^{\pm\pm}$ in the minimal LRSM are shown in Fig.~\ref{Fig:1}. The diagram (a) arises from the Yukawa coupling $f_R$ in Eq.~\eqref{eq:lag}, which leads to the decay of $H^{\pm\pm}$ into a pair of same-sign charged leptons, with the partial decay width
\begin{align}\label{yukawal}
    \Gamma(H^{\pm\pm}\to l_R^\pm l_R^\pm)\simeq \frac{f_R^2}{8\pi}m_{H^{\pm\pm}} \, .
\end{align}             
In fact, this decay mode has been used by the LHC experiments to set a lower limit on the doubly charged scalar mass: $m_{H^{\pm\pm}}> 900$ GeV~\cite{ATLAS:2022pbd}. However, this limit is valid only for {\it prompt} decay of $H^{\pm\pm}$ into a pair of isolated leptons. For $f_R\lesssim 10^{-8}$, Eq.~\eqref{yukawal} indicates that a TeV-scale $H^{\pm\pm}$ is sufficiently long-lived and decays outside the detector (see Ref.~\cite{Dev:2018tox} for a detailed discussion on this). In this case, the mass limit is weaker, at about 700 GeV, as we derive below. We first assess the required smallness of the Yukawa coupling $f_R$ for achieving long-lived behavior of $H^{\pm\pm}$.
As depicted in Fig.~\ref{Fig:2}, for a TeV mass of $H^{\pm\pm}$ and a very heavy $W_R$ mass (so that diagrams (b) and (c) in Fig.~\ref{Fig:1} are not relevant), a Yukawa coupling of $f_R \sim 10^{-20}$ is needed to attain a lifetime of $H^{\pm\pm}$ around $10^{17}$ sec [cf. Eq.~\ref{yukawal}]. 
If we impose a discrete $Z_2$ symmetry under which $\Delta_R\to - \Delta_R$ and all other fields are even, then such Yukawa couplings are forbidden in Eq.~\eqref{eq:lag}.\footnote{This also forbids the cascade decay $H^{\pm\pm}\to H^{\pm^*}H^{\pm^*}$.} In this case, the only coupling that leads to the $H^{\pm\pm}$ decay is its $SU(2)_R$ gauge coupling via which it decays to two (on/off-shell) $W_R$ bosons, which subsequently decay to hadrons and leptons [see Fig.~\ref{Fig:1}(b)].  The corresponding decay rate is given in Appendix~\ref{app:decay}.  
For $m_{H^{\pm\pm}}\ll m_{W_R}$, this scales as $v_R^{-6}$. Therefore, significantly increasing the $W_R$ mass relative to the $H^{\pm\pm}$ mass suppresses this decay rate, thus resulting in an extended lifetime for $H^{\pm\pm}$. Suppression of $W_L-W_R$ mixing is necessary to weaken the $H^{\pm\pm}\to W^\pm_L W^\pm_L$ mode, and a high $W_R$ mass achieves this objective. In Fig.~\ref{Fig:3}, we present a plot of $W_R$ mass against $H^{\pm\pm}$ mass for various lifetimes of $H^{\pm\pm}$. For instance, with $m_{H^{\pm\pm}}\simeq$ TeV and $m_{W_R}\sim 10^{10}$ GeV, the resulting lifetime of $H^{\pm \pm}$ exceeds the current age of the Universe, $\tau_U=4.34\times 10^{17}$ sec.

\section{Neutrino mass generation} \label{sec:IV}
Setting the Yukawa coupling of $\Delta_R$-triplet with leptons to zero renders the type-I seesaw~\cite{Minkowski:1977sc, Mohapatra:1979ia, Yanagida:1979as, Gell-Mann:1979vob, Glashow:1979nm} for neutrino masses inoperative. To address this, we extend the model by introducing a right-handed scalar doublet $\chi_R$ and three singlet fermions $n=(n_{1,2,3})$. The Yukawa Lagrangian is then expanded to include the following additional terms:
\begin{eqnarray}
    {\cal L}'_Y~=~ f'_a \bar{\ell}_{a,R}\chi_R n_a + \mu_{ab} n_a n_b+ {\rm H.c.}\, ,
\end{eqnarray}
where $\mu_{ab}$ is a small lepton-number-breaking parameter adjusted to fit neutrino masses.
Now, there are additional new terms in the Higgs potential as follows:
\begin{align}
 &   V'(\chi_R, \Delta_R, \Phi)~=-\mu^2_\chi \chi^\dagger_R\chi_R+ \lambda_R (\chi^\dagger_R\chi_R)^2 \nonumber \\ 
  & \qquad  +\lambda^\prime_R (\chi^\dagger_R\chi_R){\rm Tr}(\Delta^\dagger_R\Delta_R)  + \rho{''}\chi^\dagger_R\Delta^\dagger_R\Delta_R\chi_R \nonumber \\    & \qquad +\rho{'}\chi^\dagger_R\Delta_R\Delta^\dagger_R\chi_R
    +\alpha' \chi^\dagger_R\Phi^\dagger\Phi\chi_R \, .
    \label{eq:Vp}
\end{align}
We do not include the $\chi_L$ in the theory for the same reason that we do not include $\Delta_L$, i.e. because we break $D$-parity at a much higher scale. Another possible term $\mu_\chi \chi_R \Delta_R \chi_R$ is not included since it is forbidden by the discrete symmetry $\Delta_R\to -\Delta_R$ which we impose to forbid leptonic coupling of $\Delta_R$.

We choose the vacuum as follows: $\langle\chi^0_R\rangle=u_R$ with $u_R\sim $ TeV and $\langle\Delta^0_R\rangle=v_R\gg$ TeV. 
There are several implications of this extension of the model: (i) First, once $\langle\chi^0_R\rangle\neq 0$, the three neutral fermions $(\nu, \nu_R, n)$ mix with each other and via inverse seesaw~\cite{Mohapatra:1986aw, Mohapatra:1986bd} give small mass to the neutrinos\footnote{For an alternative implementation of inverse seesaw in low scale left-right models, see~\cite{Brdar:2018sbk}.}; (ii) There are new terms that break the global $SU(3)$ symmetry in addition to those mentioned earlier explaining smaller $H^{\pm\pm}$ mass, e.g. the $\rho'$ and $\rho''$ couplings in the above potential. Since $u_R\sim $ TeV, they only give TeV scale masses to the $H^{\pm\pm}$. Thus, even in the presence of the right-handed doublet Higgs field, one can have a TeV scale $H^{\pm\pm}$ with much higher values of $W_R$ mass simply by tuning down the symmetry breaking parameters $\rho_2$. The decay of $\chi^0_R\to H^{++}H^{--}$ is given by the coupling $\rho^{\prime\prime}$, which we can choose of order one. This will then help in producing a pair of long-lived doubly charged scalars at colliders. 
We will refer to the neutral scalar field $\chi^0_R$ simply as $\chi^0$. The coupling $\rho'$, on the other hand, contributes to the mass of $\chi^0$, i.e. $m_{\chi^0}^2\simeq \rho' v^2_R$. In the following collider analysis, we fix the ratio $m_{\chi^0}/v_R=10^{-6}$ in order to keep $\chi^0$ field at the TeV scale while having $v_R$ much larger, as needed to explain the $dE/dx$ excess.

\section{Collider implications} \label{sec:V}
\begin{figure*}[t!]
		\centering
  \includegraphics[width=0.8\textwidth]{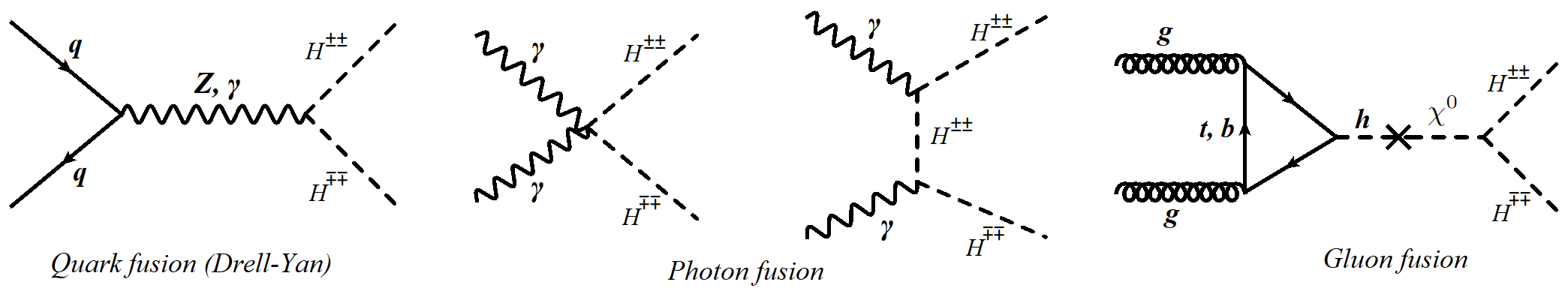}\\
\hspace{0.5cm}  (a) \hspace{5cm} (b) \hspace{5cm} (c)
	\caption{Representative Feynman diagrams for dominant production modes of $H^{\pm \pm}$ at the LHC.}
	\label{Fig:4}
\end{figure*}
Here, we investigate the phenomenological implications of long-lived doubly charged scalars in collider experiments. If these LLPs exist within the accessible energy range of the colliders, they may yield null results in prompt signal searches. However, their presence could lead to different signals like disappearing tracks and displaced vertices, depending on their lifetime, decay modes, and decay length relative to the detector.\footnote{Collider implications of neutral/charged LLPs in the context of other neutrino mass models have been widely discussed; see e.g.,~Refs.~\cite{deCampos:2005ri, Dev:2016nfr, 
Dev:2017dui, Ghosh:2018drw, Jana:2018rdf,  Abada:2018sfh, Dev:2018tox, Arbelaez:2019cmj, Das:2019fee, Jana:2020qzn, Frank:2019nid, R:2020odv, MATHUSLA:2020uve, Hirsch:2021wge, Cottin:2022nwp, Ghosh:2023tur}.} Collider experiments from LEP~\cite{OPAL:2003zpa} to Tevatron~\cite{CDF:2005hfo}, LHC~\cite{LHCb:2015ujr, CMS:2016kce, ATLAS:2022pib, ATLAS:2023zxo, ATLAS:2023esy, CMS:2023mny}, as well as dedicated detectors like MoEDAL~\cite{Acharya:2020uwc, MoEDAL:2023ost}, have searched for the signatures of stable and long-lived multiply charged particles across diverse energy scales, exploring a wide range of possibilities for their existence.  If the charged LLP lives long enough to traverse the entire detector without decaying, it leaves a distinct ionizing track in the inner detector, from which a large ionization energy gradient ($dE/dx$) signal is expected. It is noteworthy to highlight recent findings by the ATLAS collaboration, which observed an excess of 7 events characterized by a substantial energy loss gradient ($dE/dx > 2.4$ MeV g$^{-1}$ cm$^2$), with a known background of $0.7\pm0.4$ events, resulting in a local (global) significance of $3.6\sigma$ (3.3$\sigma$)~\cite{ATLAS:2022pib}. As we show below, this excess can be explained in our model setup.

In the minimal LRSM, there exist two dominant production modes for doubly charged scalars at the LHC: (a) quark-fusion process through $s$--channel $Z/\gamma$ exchange (Drell-Yan)~\cite{Gunion:1989in, FileviezPerez:2008jbu}, and (b) photon fusion processes~\cite{Babu:2016rcr}; see  Fig~\ref{Fig:4}. In the extended version for neutrino mass generation discussed above, there is an additional gluon-fusion production channel via $\chi^0$ mixing with the SM Higgs boson [cf.~Fig~\ref{Fig:4} (c)]. This latter diagram turns out to be crucial for explaining the $dE/dx$ excess. Before going into that discussion, we first derive a lower limit on the $H^{\pm\pm}$ mass following the recent ATLAS analysis~\cite{ATLAS:2023zxo} of heavy, long-lived, multi-charged particle search at the $\sqrt{s}=13 \mathrm{TeV}$ LHC with an integrated luminosity of 139 fb$^{-1}$. After incorporating our model file into the {\tt FeynRules} package~\cite{Alloul:2013bka}, we calculate the cross-section for the Drell–Yan plus photon fusion pair production mode of $H^{\pm\pm}$ at the 13 TeV LHC using the Monte Carlo event generator {\tt MadGraph5aMC@NLO}~\cite{Alwall:2014hca}. Comparing it with the observed upper limit on the same Drell–Yan plus photon fusion cross section from ATLAS, we obtain a lower limit of 700 GeV on the mass of $H^{\pm\pm}$, which is shown by the vertical dashed line in the $f_R\lesssim 10^{-8}$ region of Fig.~\ref{Fig:2}. 

However, our analysis indicates that the pair production of long-lived doubly charged scalars through the Drell-Yan and photon fusion processes yields a relatively small boost factor $\beta \lesssim 0.7$, where  $\beta(=v/c)$ denotes the velocity of the ionizing particle within the detector. On the other hand, 
 all $dE/dx$ excess events display a relatively high $\beta \approx 1$~\cite{ATLAS:2022pib}. To account for this observed excess and maintain consistency with time-of-flight measurements, a highly boosted, long-lived doubly charged particle is necessary~\cite{Giudice:2022bpq}. Therefore, the pair production of long-lived doubly charged scalars within the minimal LRSM, facilitated by quark and photon fusion processes, is unlikely to explain the observed substantial ionization energy loss gradient~\cite{ATLAS:2022pib}. However, we have identified an additional production mechanism for such long-lived doubly charged scalars within the LRSM involving the gluon fusion process [cf. Fig.~\ref{Fig:4} (c)]. Due to mixing with SM Higgs, this process resonantly produces a neutral $\chi^0$, which subsequently decays into a pair of long-lived doubly charged scalars. In this scenario, the long-lived doubly charged scalars can attain significant boosts, providing a potential explanation for the observed $dE/dx$ excess while maintaining consistency with time-of-flight measurements.

One might wonder whether any of the neutral scalar fields in the minimal LRSM can also qualify to act as the parent resonance instead of $\chi^0$.  
As previously discussed, the masses of the bidoublet scalars ($m_{H_1^0, A_1^0}$) are constrained to be above 15 TeV or so to adhere to the FCNC constraints~\cite{Zhang:2007da}, so they cannot be resonantly produced at the LHC. On the other hand, it is possible to have the mass scale of $m_{H_3^0}$  significantly lower than the $v_R$ scale, depending on the tunable quartic coupling $\rho_{1}$ [cf.~Eq.~\eqref{eq:H3}]~\cite{Dev:2016nfr, Dev:2017dui}. In this context, it is crucial to assess the radiative stability of such a configuration. In the SM, a lower bound of 5 GeV exists for the SM Higgs boson mass~\cite{Linde:1975sw, Weinberg:1976pe}, considering only contributions from gauge bosons and neglecting fermionic contributions in the Coleman-Weinberg effective potential~\cite{Coleman:1973jx}. Given that we are setting the Yukawa coupling $f_R$ to zero in our scenario, a lower bound on $m_{H_3^0}$ emerges when considering contributions solely from right-handed gauge bosons in the Coleman-Weinberg effective potential. Analogous scenarios, as demonstrated in Ref.~\cite{Holthausen:2009uc}, establish a lower bound of 900 GeV when considering a neutral scalar originating from an $SU(2)_R$-doublet. Consequently, as long as $m_{H_3^0}\gtrsim$ TeV, it satisfies these theoretical constraints. However, the cubic coupling $H_3^0 H^{++}H^{--}$ is proportional to $2\sqrt{2}(\rho_1+2\rho_2)v_R$, while the cubic coupling $H_3^0 hh$ is proportional to $\alpha_1 v_R/\sqrt{2}$~\cite{Dev:2016dja}. Setting the $v_R$ scale to a high-value results in small values for $\rho_{1,2}$ to maintain the masses of $m_{H_3^0, H^{\pm\pm}}$ at TeV scale. The important point here is that despite the cubic coupling's dependence on $v_R$, this configuration favors the dominance of the di-Higgs branching ratio mode in the decay of $H_3^0$, making the $H^{++}H^{--}$ branching mode significantly small. This is attributed to the mass-coupling relations, specifically expressed as 
\begin{align}
    \frac{{\rm Br} (H_3^0 \to hh)}{{\rm Br} (H_3^0 \to H^{++}H^{--})}=\frac{\sin^2\vartheta(\frac{v_R}{\kappa})^2 }{ 4(1+\frac{2\rho_2}{\rho_1})^2} \, ,
    \label{eq:BR}
\end{align}
where $\vartheta$ represents the mixing between the SM Higgs $h$ and $H_3^0$. Because the same mixing parameter is responsible for the $H_3^0$ production via gluon fusion, it cannot be arbitrarily small. In fact, to reproduce the observed value of the cross-section for the $dE/dx$ excess, we need $\sin\theta\gtrsim 0.1$ (see below), which implies from Eq.~\eqref{eq:BR} that the $H_3^0 \to hh$ mode will always be dominant. Due to these interconnected mass-coupling relationships, $H_3^0$ does not qualify as a suitable parent particle responsible for the observed excess in $dE/dx$.

As we discussed in Section~\ref{sec:IV}, within the inverse seesaw extension of the LRSM, we have the neutral component of the doublet scalar $\chi^0$, which can also mix with the SM Higgs, characterized by the mixing angle $\theta$, via the $\alpha'$ term in the potential~\eqref{eq:Vp}.
The intriguing outcome of this mixing is the resonant production of $\chi^0$ at the LHC through the gluon fusion process [cf.~Fig.~\ref{Fig:4} (c)]. The magnitude of the mixing angle $\theta$ is restricted from Higgs signal strength data obtained at the LHC, which currently permits $\sin~\theta<0.33$ from a recent CMS analysis~\cite{CMS:2022dwd}.\footnote{A slightly stronger bound of $\sin~\theta<0.26$ is obtained from the ATLAS analysis~\cite{ATLAS:2022vkf}.} In our analysis, we assume that the VEV $\langle\chi^0_R\rangle$ is significantly smaller than the right-handed symmetry-breaking scale $v_R$ and place it in the TeV range. As explained above, this ensures that the cubic coupling $\chi^0 H^{++}H^{--}$ remains unsuppressed, allowing $\chi_R$ to predominantly decay into $H^{++}H^{--}$, and not to $hh$.
Consequently, following resonant production at the LHC, $\chi^0$ primarily undergoes decay into $H^{++}H^{--}$. In this scenario, the doubly charged scalars exhibit a substantial boost, as required to explain the observed $dE/dx$ excess while maintaining agreement with time-of-flight measurements.

\begin{figure}[t!]
		\centering
  \includegraphics[width=0.49\textwidth]{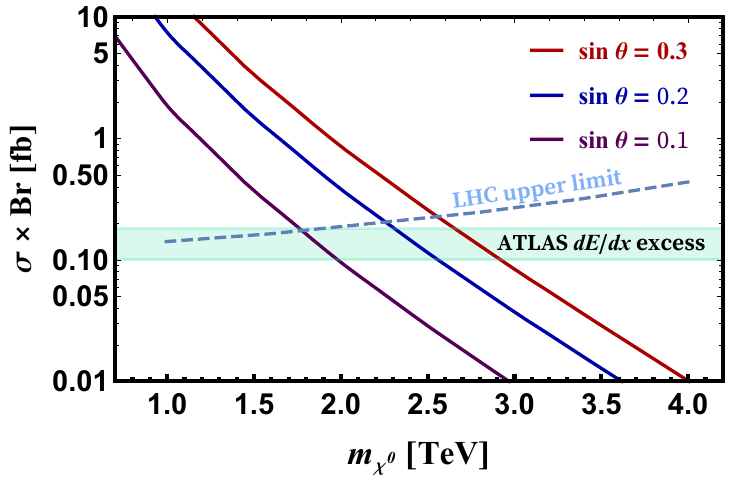} 
	\caption{The cross-section times branching ratio for resonant production of $\chi^0$ and decay into a pair of long-lived doubly charged scalars in the LRSM at the $\sqrt s=13$ TeV LHC  via gluon fusion process as a function of the mass of $\chi^0$ for different values of mixing with the SM Higgs. The green band represents the preferred range, which explains the observed $dE/dx$ excess at 95\% CL~\cite{ATLAS:2022pib, Giudice:2022bpq}. The blue dashed curve represents the experimental upper limit on the doubly charged scalar pair-production cross-section, assuming $m_{\chi^0}=2m_{H^{\pm\pm}}$.  
 }
	\label{Fig:6}
\end{figure} 

Fig.~\ref{Fig:6} illustrates the cross-section, $ \sigma (pp \to \chi^0) \times \mathrm{Br} (\chi^0 \to H^{++}H^{--})$, at the $\sqrt s=13$ TeV LHC as a function of the mass of $\chi^0$ for various values of the mixing angle $\theta$. The green band represents the preferred range of cross-section (0.10--0.19 fb), which explains the observed $dE/dx$ excess at 95$\%$ CL~\cite{ATLAS:2022pib, Giudice:2022bpq}. Notably, for $\chi^0$ masses exceeding approximately 3 TeV, the production cross-section experiences a sharp decline, rendering it insufficient to account for the observed excess, even for the largest possible value of $\sin\theta_{\rm max}\simeq 0.3$ (corresponding to the red curve in Fig.~\ref{Fig:6}). Also shown in Fig.~\ref{Fig:6} (blue dashed curve) is the 95\% CL upper bound on the cross section $\sigma(pp\to H^{++}H^{--})$ from the recent ATLAS analysis~\cite{ATLAS:2023zxo} for a fixed ratio of $m_{\chi^0}/m_{H^{\pm\pm}}=2$. It is clear that the $dE/dx$ preferred region from the previous ATLAS analysis~\cite{ATLAS:2022pib} is still consistent with the latest ATLAS analysis~\cite{ATLAS:2023zxo} in our model for $m_{\chi^0}>2$ TeV.  

\begin{figure}[t!]
		\centering
		\includegraphics[width=0.49\textwidth]{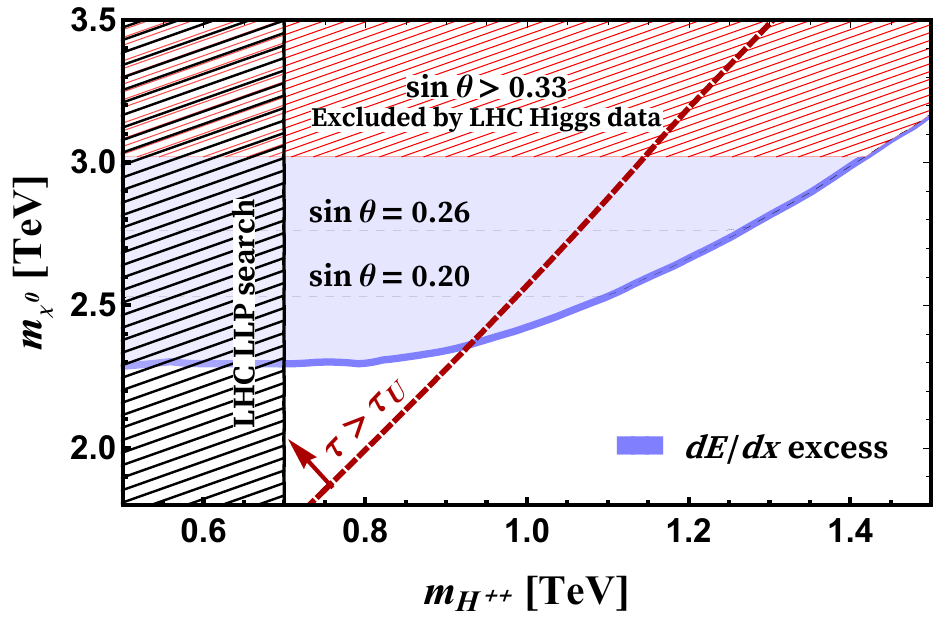}
	\caption{Model parameter space in $m_{H^{++}}$ vs $m_{\chi^{0}}$ plane to explain the $dE/dx$
 excess at 95\% CL (blue-shaded region). Current LHC data from LLP searches~\cite{ATLAS:2023zxo} excludes the black-hatched region. The red-hatched region is excluded by the LHC data on the Higgs signal strength observables~\cite{CMS:2022dwd}. The red dashed line corresponds to the lifetime $\tau=\tau_U=4.34\times 10^{17}$ sec. 
 The two black dashed horizontal lines correspond to mixing values $\sin~\theta = 0.20$ and 0.26.   
 }
	\label{Fig:7}
\end{figure} 

Fig.~\ref{Fig:7} depicts a comprehensive view of the allowed parameter space in the $m_{H^{++}}$ versus $m_{\chi^{0}}$ plane to explain the $dE/dx$ excess. The blue-shaded region represents the 95\% CL preferred region, taking into account the number of events necessary for reproducing the best-fit $dE/dx$ excess signal~\cite{Giudice:2022bpq}:
\begin{align}
    & N_{\mathrm{ev}}\left(p p \rightarrow \chi^0 \rightarrow H^{++}H^{--}\right)= \mathcal{L} \, \epsilon \, \sigma (pp \to \chi^0) \nonumber \\
    & \qquad \qquad \qquad \qquad \qquad 
    \times \mathrm{Br} (\chi^0 \to H^{++}H^{--}) \, ,
\end{align}
where ${\cal L}$ is the integrated luminosity (taken here as 139 fb$^{-1}$) and $\epsilon$ is the signal efficiency factor (taken here to be 20\%). Following Ref.~\cite{Giudice:2022bpq}, we take $N_{\mathrm{ev}}=5$ to derive our best-fit region in Fig.~\ref{Fig:7}. We ensure that these events originate from anomalously large ionizing tracks attributed to rapidly moving ($\beta\approx 1$), long-lived doubly charged scalars $H^{\pm \pm}$ and are consistent with the observed $m_{dE/dx}$ distribution~\cite{Giudice:2022bpq}. Furthermore, we fix ${v_R}/{m_{\chi^0}}=10^6$ (thus inherently fixing the $W_R$ mass) to optimize the scenario, enabling us to achieve a doubly charged scalar lifetime $\tau > \tau_U$ within the parameter space of our interest, as shown by the red dashed line.  Notably, within the mass range of $\chi^0$ from 2.2 TeV to 3.1 TeV and $m_{H^{++}}$ from 700 GeV up to $m_{\chi^0}/2$, a considerable parameter space exists, yielding a substantial ionization energy loss gradient capable of explaining the $dE/dx$ excess. 

Note that reducing the $v_R$ value, and consequently the $W_R$ mass, shortens the decay lifetime of the $H^{\pm \pm}$ particles, leading to shorter decay lengths [cf.~Fig.~\ref{Fig:3}]. To address the $dE/dx$ excess, it is crucial to ensure that the proper decay lengths of $H^{\pm \pm}$ particles exceed $c\tau \gtrsim 1$ mm ($\tau \gtrsim10^{-11}$ sec) to avoid prompt decay signatures. For instance, for a TeV mass of $H^{\pm \pm}$ particles, the $W_R$ mass should be roughly heavier than 20 TeV, requiring $v_R$ to be greater than 43 TeV. While this renders the $H^{\pm \pm}$ lifetime too short for catalyzed fusion, it still induces significant ionization energy loss, elucidating the observed $dE/dx$ excess as illustrated in Fig.~\ref{Fig:7}.

Outside the blue region, either the production cross-section proves inadequate to explain the observed events, or the $H^{\pm \pm}$ particles fail to attain the requisite boost. It was shown in Ref.~\cite{Giudice:2022bpq} that if we also take the $p_T$ distribution into account, combined with the $m_{dE/dx}$, the optimal range for the mass of the parent particle $\chi^0$ is found to be above $3$ TeV. However, in order to achieve a sufficient cross-section above $3$ TeV in our Higgs mixing scenario, a larger mixing angle $\theta$ is required, which is excluded by the current Higgs signal strength data from the LHC, as shown by the red hatched region in Fig.~\ref{Fig:7}. Here we have used the CMS analysis~\cite{CMS:2022dwd} which gives $\sin\theta<0.33$; using the ATLAS analysis~\cite{ATLAS:2022vkf} would give $\sin\theta<0.26$ (also indicated in Fig.~\ref{Fig:7}).  The presented region in our analysis specifically yields a favorable fit to the $m_{dE/dX}$ distribution. Notably, it requires the Higgs mixing angle $\sin\theta$ to be above 0.15, which can be conclusively tested in future precision Higgs data at the HL-LHC or at future colliders~\cite{deBlas:2019rxi}, should the $dE/dx$ excess persist at LHC Run-3.   

\section{$H^{--}$--catalyzed fusion mechanism} \label{sec:VI}
Long-lived doubly charged particles have the potential to catalyze the fusion of light nuclei, presenting a significant possibility for applications in energy production~\cite{Akhmedov:2021qmr}. Specifically, the negatively-charged $H^{--}$ can establish atomic bound systems with light element nuclei, including deuterium, tritium, or helium. Such atomic systems will be characterized by very small Bohr radii, which will essentially eliminate the necessity for the light nuclei within them to overcome the Coulomb barrier in order to undergo fusion. There are several articles in the literature discussing the potential of generating energy through nuclear fusion with hypothetical heavy, long-lived or stable singly-charged~\cite{Zeldovich:1958, Ioffe:1979tv, Rafelski:1989pz, Hamaguchi:2006vp, Hamaguchi:2012px} or fractionally charged~\cite{Zweig:1978sb} particles. However, these approaches encountered catalytic poisoning challenges, making them unsuitable for practical energy production. Some studies~\cite{Ioffe:1979tv} proposed reactivating catalyst particles, but this would require neutron beams significantly exceeding the current capabilities of nuclear reactors. 

This study focuses on catalyzing light nuclei fusion with doubly charged $H^{--}$ particles within the LRSM, highlighting its potential as a feasible energy source. Catalytic poisoning, in this case, has a very small probability, and each $H^{--}$ can catalyze up to $3.5\times 10^9$ deuterium fusion cycles, producing $\sim 7\times 10^4$\, TeV of energy before being knocked out of the catalytic process due to sticking to the produced $^6$Li nuclei. Unstable $H^{--}$ have to be produced at accelerators; the catalytic
process is then energetically favorable for $H^{--}$ lifetimes above $9.5\times 10^{11}$\,sec, but in general, requires reactivation and reuse of the catalyst particles. For stable or practically stable $H^{--}$ ($\tau > \tau_U = 4.34\times 10^{17}$\,sec), there should exist a terrestrial population of relic $H^{--}$ bound to nuclei and thus forming exotic atoms; they can be extracted from continental crust or marine sediments at low energy cost instead of being produced at accelerators. No reactivation of the catalyst particles is necessary in this case. 

In Fig.~\ref{Fig:3}, we show by red dot-dashed and dashed lines the two important benchmark lifetimes ($\tau=9.5\times 10^{11}$\,sec and $\tau = 4.34\times 10^{17}$\,sec, respectively) relevant for the catalyzed fusion mechanism in the minimal LRSM. These correspond to the minimum $SU(2)_R$-breaking scales of $v_R\gtrsim 4\times 10^5$ TeV and $3\times 10^6$ TeV, respectively. Thus, we find that the doubly charged scalars in the minimal LRSM are a promising candidate for the catalyzed fusion mechanism. 

We note that the available experimental searches of exotic atoms on the earth allow for the existence of stable doubly charged particles in abundances suitable for their practical use for catalyzed fusion, provided that their mass exceeds 1.2\, TeV~\cite{Smith:1982qu}; see Ref.~\cite{Akhmedov:2021qmr} for more details.
Constraints on massive charged relics have also been derived from astrophysical observations, such as abundances of old white dwarfs~\cite{Fedderke:2019jur}. These constraints, however, are rather weak for moderately light charged relics with masses of interest to us ($10^3$ -- $10^5$\, GeV) and are model-dependent~\cite{Fedderke:2019jur, Chuzhoy:2008zy}; they do not challenge the prospects of using catalyzed fusion for energy production.

\section{Cosmological constraints}

Fusion of light nuclei catalyzed by doubly negatively charged scalars may also play an important role in big-bang nucleosynthesis (BBN). If present in considerable amounts, they could catalyze the production of light elements at the BBN epoch, which in particular could lead to the overproduction of Li, Be, and B. Consistency with observations requires the abundance of the doubly negatively charged scalars to be below $\sim 10^{-9}$ compared to hydrogen~\cite{AP}. We note that this constraint does not jeopardize the prospects of using catalyzed fusion for energy production~\cite{Akhmedov:2021qmr}.

There are cosmological constraints on the abundances of stable charged particles coming from upper bounds on their annihilation cross-sections, including unitarity bounds (see e.g.  Ref.~\cite{Berger:2008ti}). These upper bounds lead to lower limits on the expected abundances of stable charged relics. Consistency with the upper bounds coming from direct terrestrial searches for exotic atoms formed by charged relics then allows one to exclude the existence of stable charged relics in a significant
range of their masses, including the TeV scale, which is of interest to us.  However, these constraints apply only to thermal relics and can readily be avoided if the masses of charged relics are significantly larger than the
temperature of reheating after inflation, as this would strongly suppress their abundance compared to the thermal one due to entropy production. It should be noted that reheat temperatures as low as a few MeV are consistent with observations~\cite{Hannestad:2004px}. For such low reheat temperatures, TeV-scale stable charged relics are allowed~\cite{Kudo:2001ie}.  

More detailed analysis of cosmological implications of the existence of very long-lived or stable doubly charged particles is beyond the scope of the present Letter.
\section{Conclusions} \label{sec:VII}
Doubly charged particles frequently emerge in BSM scenarios. However, they do not typically qualify as long-lived particles. Here we have shown that the doubly charged scalars originating from the $SU(2)_R$-triplet Higgs field within the Left-Right Symmetric Model can be long-lived. By virtue of a concealed symmetry in the scalar potential, the mass of these doubly charged particles (denoted here by $H^{\pm\pm}$) can be significantly lower than the $SU(2)_R$-breaking scale at the tree level, rendering them sufficiently long-lived (even cosmologically stable), while adhering to all prevailing theoretical and experimental constraints. 
This unique feature positions these $H^{\pm\pm}$ particles as ideal candidates for catalyzed fusion processes in light nuclei, offering promising prospects for energy production. Incorporating neutrino mass through the inverse seesaw mechanism and simultaneously allowing the existence of long-lived doubly charged scalars enables us to provide an explanation for the recently observed ionization energy loss ($dE/dx$) excess in the LHC data. Consequently, this model may carry substantial implications for heavy stable charged particle searches in future collider experiments, and can be independently tested in future precision Higgs data. 

\section*{Acknowledgments}
EA is grateful to Goran Senjanovi\'{c} for an illuminating email correspondence. BD thanks Yongchao Zhang for useful discussion, and acknowledges the local hospitality of MPIK where this work was originally conceived. BD and SJ acknowledge the Fermilab Theoretical Physics Department, where part of this work was done, for their warm hospitality. BD and SJ also wish to thank the Center for Theoretical Underground Physics and Related Areas (CETUP*) and the Institute for Underground Science at SURF for their hospitality and for providing a stimulating environment. The work of BD was partly supported by the U.S. Department of Energy under grant No. DE-SC 0017987.

\begin{widetext}
\appendix
\section{Scalar Sector of the LRSM} \label{sec:app}
In the minimal LRSM, the most general renormalizable scalar potential, invariant under parity, is given by~\cite{Deshpande:1990ip, Maiezza:2016ybz, Dev:2018xya, Chauhan:2019fji}
\begin{align}
V(\Phi, \Delta_L, \Delta_R) =&
-\mu^2_1{\rm Tr} (\Phi^\dagger\Phi)-\mu^2_2\left[{\rm Tr} (\Phi^\dagger\tilde{\Phi})+{\rm Tr} (\Phi\tilde{\Phi}^\dagger)\right]   -\mu^2_3\left[{\rm Tr} (\Delta^\dagger_L\Delta_L) + {\rm Tr} (\Delta^\dagger_R\Delta_R) \right] \nonumber \\
& +\lambda_1\left[{\rm Tr} (\Phi^\dagger\Phi)\right]^2+\lambda_2\left\{ \left[ {\rm Tr} (\Phi^\dagger\tilde{\Phi})\right]^2  +\left[{\rm Tr} (\Phi\tilde{\Phi}^\dagger)\right]^2\right\}  
\nonumber \\
&
+\lambda_3 {\rm Tr} (\Phi^\dagger\tilde{\Phi}){\rm Tr} (\Phi\tilde{\Phi}^\dagger)  +\lambda_4 {\rm Tr} (\Phi^\dagger\Phi)
    \left[{\rm Tr} (\Phi\tilde{\Phi}^\dagger)+{\rm Tr} (\Phi^\dagger\tilde{\Phi})\right] \nonumber \\
    & +\rho_1 \left\{\left[{\rm Tr} (\Delta^\dagger_L \Delta_L)\right]^2+\left[{\rm Tr} (\Delta^\dagger_R \Delta_R)\right]^2\right\} +   \rho_2\left[{\rm Tr} (\Delta_L \Delta_L) {\rm Tr} (\Delta^\dagger_L \Delta^\dagger_L)+{\rm Tr} (\Delta_R \Delta_R) {\rm Tr} (\Delta^\dagger_R \Delta^\dagger_R) \right]\nonumber \\
    & + \rho_3{\rm Tr}(\Delta_L^\dag \Delta_L){\rm Tr}(\Delta_R^\dag \Delta_R) 
    + \rho_4 \left[{\rm Tr} (\Delta_L \Delta_L) {\rm Tr} (\Delta^\dagger_R \Delta^\dagger_R)+{\rm Tr} (\Delta_R \Delta_R) {\rm Tr} (\Delta^\dagger_L \Delta^\dagger_L) \right] \nonumber \\
    & +\alpha_1 {\rm Tr}(\Phi^\dag \Phi)\left[ {\rm Tr} (\Delta^\dagger_L \Delta_L)+{\rm Tr} (\Delta^\dagger_R \Delta_R)\right]+ \left\{\alpha_2e^{i\delta_2}\left[{\rm Tr}(\Phi^\dagger\tilde{\Phi }) {\rm Tr} (\Delta^\dagger_L \Delta_L)+ {\rm Tr}(\Phi\tilde{\Phi }^\dag) {\rm Tr} (\Delta^\dagger_R \Delta_R)\right]+  {\rm H.c.}\right\}\nonumber \\
    & +\alpha_3\left[{\rm Tr}(\Phi\Phi^\dag \Delta_L\Delta_L^\dag)+{\rm Tr}(\Phi^\dagger\Phi \Delta_R\Delta_R^\dag)\right]
    +\beta_1\left[{\rm Tr}(\Phi \Delta_R\Phi^\dag\Delta_L^\dag)+{\rm Tr}(\Phi^\dag\Delta_L\Phi\Delta_R^\dag) \right] \nonumber \\
    &+\beta_2\left[{\rm Tr}(\tilde{\Phi} \Delta_R\Phi^\dag\Delta_L^\dag)+{\rm Tr}(\tilde{\Phi}^\dag\Delta_L\Phi\Delta_R^\dag) \right]
    + \beta_3\left[{\rm Tr}(\Phi \Delta_R\tilde{\Phi}^\dag\Delta_L^\dag)+{\rm Tr}(\Phi^\dag\Delta_L\tilde{\Phi}\Delta_R^\dag) \right]
    .
    \label{eq:pot}
\end{align}
Due to the left-right symmetry, all 18 parameters $\mu^2_{1,2,3}$, $\lambda_{1,2,3,4}$, $\rho_{1,2,3,4}$, $\alpha_{1,2,3}$ and $\beta_{1,2,3}$ are real, and the only phase associated with $\alpha_2$ is explicitly shown. Minimizing the potential with respect to the VEVs given in Eq.~\eqref{eq:vev}, we can in fact remove many of these parameters. 

Following the convention in Ref.~\cite{Zhang:2007da}, the VEVs of the scalar fields are given by
\begin{eqnarray}
\langle \Phi \rangle = \left(\begin{array}{cc} \kappa &0\\0 & \kappa'e^{i\delta}\end{array}\right) \, , \quad 
\langle \Delta_L \rangle = \left(\begin{array}{cc} 0&0\\v_Le^{i\theta} & 0 \end{array}\right) \, , \quad 
\langle \Delta_R \rangle = \left(\begin{array}{cc} 0&0\\v_R & 0 \end{array}\right) \, .
\label{eq:vev}
\end{eqnarray}In terms of the scalar fields, expanding them around the corresponding VEVs, we get 20 degrees of freedom in total, which can be separated into the neutral, singly-charged and doubly charged components, as follows: 
\begin{align}
   {\textrm {Neutral}}: & \left\{\Re({\phi_1^0}), \Re(\phi_2^0), \Re(\Delta_L^0), \Re(\Delta_R^0),\Im({\phi_1^0}), \Im(\phi_2^0), \Im(\Delta_L^0), \Im(\Delta_R^0)\right\} \, . \nonumber \\
   {\textrm {Singly-charged}}: & \left\{\phi_1^\pm, \phi_2^\pm, \Delta_L^\pm, \Delta_R^\pm \right\}\, . \nonumber \\
 {\textrm {doubly charged}}: & \left\{\Delta_L^{\pm\pm},\Delta_R^{\pm\pm} \right\} \, .
\end{align}
In the physical mass basis, two neutral components and two pairs of singly-charged components are identified as the Goldstone modes which are responsible for the masses of the two neutral ($Z,Z_R$) and two pairs of charged ($W^\pm, W_R^\pm$) gauge bosons, respectively. Thus, we are left with 14 physical scalar fields. 

Due to the large number of parameters involved, the general expressions for the mass eigenvalues are complicated. However, a few simplifying assumptions can be made here. First, the $\Delta_L$ VEV is restricted to be small from electroweak precision data, and in particular, from $\rho$-parameter constraint: $v_L\lesssim 2$ GeV~\cite{Hollik:1986gg, Akeroyd:2005gt}. In practice, since $v_L$ contributes to the neutrino masses, it cannot be much larger than the eV scale (without introducing too much fine-tuning). Therefore, it is a natural choice to ignore $v_L$ (compared to the VEVs $\kappa$ and $v_R$) and the quartic couplings $\beta_i$ in Eq.~\eqref{eq:pot} while deriving the mass formulas. Similarly, in light of the third-generation fermion masses, we expect $\xi=\kappa'/\kappa\lesssim m_b/m_t\simeq 0.03$ (again unless we allow for fine-tuning). Finally, given the experimental lower limit of a few TeV on the $W_R$ mass~\cite{CMS:2021dzb, ATLAS:2023cjo}, we have $\epsilon=\kappa/v_R\ll 1$. Furuthermore, the CP observables require that the phase $\delta\ll 1$~\cite{Zhang:2007da}. 

Taking all this into account, we can do a perturbative expansion in the small parameters $\epsilon, \xi$ and $\alpha$ and only keep the terms up to quadratic order. For the neutral scalar sector, the eigenvalues for the six physical mass eigenstates are thus given by~\cite{Maiezza:2016ybz} 
\begin{align}
    m_h^2 = & \left(4\lambda_1-\frac{\alpha_1^2}{\rho_1}\right)\kappa^2 \, , \label{eq:h} \\
    m_{H_1^0}^2 = & \alpha_3(1+2\xi^2)v_R^2+4\left(2\lambda_2+\lambda_3+\frac{4\alpha_2^2}{\alpha_3-4\rho_1}\right)\kappa^2 \, , \\
    m_{H_2^0}^2 = & (\rho_3-2\rho_1)v_R^2 \, , \\
    m_{H_3^0}^2 = & 4\rho_1v_R^2+\left(\frac{\alpha_1^2}{\rho_1}-\frac{16\alpha_2^2}{\alpha_3-4\rho_1}\right)\kappa^2 \, , \\
    m_{A_1^0}^2 = & \alpha_3(1+2\xi^2)v_R^2+4(\lambda_3-2\lambda_2)\kappa^2 \, , \\
    m_{A_2^0}^2 = & (\rho_3-2\rho_1)v_R^2 \, .
\end{align}
Here $h$ can be readily identified as the SM-like Higgs boson, with its mass being proportional to the VEV $\kappa$ and independent of $v_R$, i.e.~$m_h\simeq 125$ GeV. $H_2^0$ and $A_2^0$ are just the mass eigenstates of $\Re(\Delta_L^0)$ and $\Im(\Delta_L^0)$ respectively, whereas $H_3^0$ is mostly composed of $\Re(\Delta_R^0)$ with a small mixing with $h$. 

In the singly-charged sector, the mass eigenvalues are 
\begin{align}
m_{H_1^\pm}^2 = & \alpha_3(1+2\xi^2)v_R^2+\frac{1}{2}\alpha_3\kappa^2  \, , \\
m_{H_2^\pm}^2 = & (\rho_3-2\rho_1)v_R^2+\frac{1}{2}\alpha_3\kappa^2 \, .
\end{align}
The doubly charged mass eigenvalues are given by 
\begin{align}
    m_{H_1^{\pm\pm}}^2 = & (\rho_3-2\rho_1)v_R^2+\alpha_3\kappa^2 \, , \\
    m_{H_2^{\pm\pm}}^2 = & 4\rho_2v_R^2+\alpha_3 \kappa^2 \, ,
\end{align}
where $H_{1,2}^{\pm\pm}$ are essentially the mass eigenstates of $\Delta_{L,R}^{\pm\pm}$ with no mixing between them up to ${\cal O}(\epsilon^2)$. 

It is easy to see from the above equations for Higgs boson masses that in the $D$-parity conserving version of the model, the $H^{\pm}_1$ is lighter than $H^{\pm\pm}_1$, because $\alpha_3$ has to be positive (since the heavy bidoublet mass is proportional to $\alpha_3$).  Therefore $H^{\pm\pm}_1$ decays quite fast as $H^{\pm\pm} _1 \to H^\pm_1+ W^{\pm *}$ (via the gauge coupling and independent of $v_L$) and does not qualify as a long-lived boson. This leaves $H^{\pm\pm}_2$ (denoted in the text by $H^{\pm\pm}$) as the only long-lived doubly charged particle in the theory.

In the $D$-parity broken version, $\Delta_L^{++}$ and $\Delta^{0,+}_L$ have  additional contribution to their masses proportional to the $D$-parity breaking scale $m_P$ and become  very heavy since $m_P \gg v_R$. In the low-energy version of this theory, therefore the $\Delta_L$ field can be integrated out and  we can drop mass eigenstates $H_2^0, A_2^0, H_1^{\pm\pm}, H^\pm_1$ from our consideration. The eigenvalues of the remaining eight mass eigenstates are given in Eqs.~\eqref{eq:H}-\eqref{eq:Hpp}, where we have kept only the leading-order terms, and have renamed the charged scalars ($H^\pm \equiv H_1^\pm$, $H^{\pm\pm}\equiv H_2^{\pm\pm}$) by dropping the subscripts for brevity.  

\section{$H^{++}$ decays via right-handed gauge bosons}
\label{app:decay}
The partial width for the decay of $H^{++}$ into the right-handed gauge bosons [cf.~Fig.~\ref{Fig:1}(b)] is given by~\cite{Dev:2018tox}
\begin{align}
& \Gamma(H^{++}\to W_R^{+(*)}W_R^{+(*)}\to f\bar{f}'f''\bar{f}''') \simeq  \frac{m_{H^{++}}^3}{16\pi^3 v_R^2} \int_0^{m_{H^{++}}^2} dp \int_0^{(m_{H^{++}}-\sqrt{p})^2} dq\nonumber \\
& \times \lambda^{1/2}(p,q,m_{H^{++}}^2)\left[\lambda(p,q,m_{H^{++}}^2)+\frac{12pq}{m_{H^{++}}^4}\right]\left[\frac{m_{W_R}\Gamma_{W_R}}{(p-m_{W_R}^2)^2+m_{W_R}^2\Gamma_{W_R}^2}\right]\left[\frac{m_{W_R}\Gamma_{W_R}}{(q-m_{W_R}^2)^2+m_{W_R}^2\Gamma_{W_R}^2}\right] \, ,
\end{align}
where $\Gamma_{W_R}\simeq \frac{g_R^2}{4\pi}m_{W_R}$ is the total decay width of the $W_R$ boson and $\lambda(x,y,z)\equiv \left(1-\frac{x}{z}-\frac{y}{z}\right)^2-\frac{4xy}{z^2}$.  Since $m_{W_R}^2=\frac{1}{2}g_R^2v_R^2$, in the limit of $m_{H^{++}}\ll m_{W_R}$, the decay rate scales as $m_{H^{++}}^7/v_R^6$, as expected for a four-body decay. 

Finally, the one-loop diagram in Fig.~\ref{Fig:1}(c) is a higher-order correction to the tree-level diagram in Fig.~\ref{Fig:1}(a), and therefore, is expected to be sub-dominant. The way to see this is as follows: It crucially depends on the Majorana mass of the right-handed neutrino in the loop. Therefore, in the limit of $f_R=0$ which implies $m_N=0$, this diagram must vanish. A detailed computation of the loop diagram (following Ref.~\cite{Zeleny-Mora:2021tym}) indeed confirms this to be true. Since we are working in this limit, we do not consider this diagram in our lifetime analysis.  
\end{widetext}

\bibliographystyle{utcaps_mod}
\bibliography{ref.bib}
\end{document}